\documentclass[runningheads]{llncs}
\usepackage{graphicx}
\usepackage{textcomp}
\usepackage{comment}
\usepackage{todonotes}
\usepackage{color}
\usepackage{array}
\usepackage{multirow}
\usepackage{hyperref}

\newcolumntype{C}[1]{>{\centering\let\newline\\\arraybackslash\hspace{0pt}}m{#1}}
\newcolumntype{L}[1]{>{\let\newline\\\arraybackslash\hspace{0pt}}m{#1}}
\usepackage{amsmath}

\DeclareMathOperator*{\argmin}{arg\,\min}
\begin{document}
\title{AI-enabled Assessment of Cardiac Systolic and Diastolic Function from Echocardiography}

\author{Esther Puyol-Ant\'on \inst{1}  
Bram Ruijsink \inst{1, 4}
Baldeep S. Sidhu  \inst{1, 2}
Justin Gould \inst{1, 2}
Bradley Porter \inst{1, 2}
Mark K. Elliott \inst{1, 2}
Vishal Mehta\inst{1, 2}
Haotian Gu \inst{3}
Christopher A. Rinaldi \inst{1, 2}
Martin Cowie \inst{5}
Phil Chowienczyk \inst{1, 3}
Reza Razavi \inst{1, 2} \and
Andrew P. King \inst{1}}
\authorrunning{E Puyol-Ant\'on et al.}   
\titlerunning{Echocardiography AI-enabled Assessment of Cardiac Function}
\institute{School of Biomedical Engineering \& Imaging Sciences, King\textquotesingle s College London, UK \and Guy’s and St Thomas\textquotesingle{} Hospital, London, UK. \and British Heart Foundation Centre, King’s College London. London, UK \and
Division of Heart and Lungs, University Medical Center Utrecht, Netherlands \and Royal Brompton Hospital (Guy’s and St Thomas\textquotesingle{} NHS Foundation Trust), UK}


\maketitle              

\begin{abstract} 
Left ventricular (LV) function is an important factor in terms of patient management, outcome, and long-term survival of patients with heart disease. The most recently published clinical guidelines for heart failure recognise that over reliance on only one measure of cardiac function (LV ejection fraction) as a diagnostic and treatment stratification biomarker is suboptimal. Recent advances in AI-based echocardiography analysis have shown excellent results on automated estimation of LV volumes and LV ejection fraction. However, from time-varying 2-D echocardiography acquisition, a richer description of cardiac function can be obtained by estimating functional biomarkers from the complete cardiac cycle. In this work we propose for the first time an AI approach for deriving advanced biomarkers of systolic and diastolic LV function from 2-D echocardiography based on segmentations of the full cardiac cycle. These biomarkers will allow clinicians to obtain a much richer picture of the heart in health and disease. The AI model is based on the 'nn-Unet' framework and was trained and tested using four different databases. Results show excellent agreement between manual and automated analysis and showcase the potential of the advanced systolic and diastolic biomarkers for patient stratification. Finally, for a subset of 50 cases, we perform a correlation analysis between clinical biomarkers derived from echocardiography and cardiac magnetic resonance and we show a very strong relationship between the two modalities.


\end{abstract}

\section{Introduction}
There has been much interest in the recent literature in using artificial intelligence (AI), or more specifically deep learning (DL), to automate the estimation of cardiac functional biomarkers from 2-D echocardiography images. For example, \cite{leclerc2019deep} evaluated a number of encoder-decoder convolutional neural network (CNN) architectures for multi-structure segmentation from apical 4- and 2-chamber echocardiography images. Their models segmented the left ventricle (LV) endo/epicardium and left atrium (LA) and used these segmentations to estimate LV ejection fraction using the public CAMUS dataset. Similarly, \cite{ouyang2020video} proposed a CNN-based method to segment the LV from  apical 4-chamber echocardiography cine acquisitions. This paper also described the publicly available EchoNet Dynamic dataset, which contains ground truth segmentations of the end diastole (ED) and end systole (ES) frames of the videos.
An alternative approach to estimating functional biomarkers is to bypass the segmentation step and attempt to estimate the biomarkers directly from the images. This approach to LV ejection fraction estimation was taken by \cite{asch2019automated} and \cite{ouyang2020video}. 
In \cite{zhang2018fully} a more complete pipeline for echocardiography image analysis was presented, incorporating view classification, CNN-based segmentation and estimation of LV/LA volumes, LV mass, LV ejection fraction and global longitudinal LV strain. The authors also demonstrated a CNN for disease classification into 3 diagnostic classes. Another clinical pipeline featuring extensive evaluation on multiple external datasets was presented in \cite{Tromp2022}. This work featured estimation of LV and LA volumes at ED and ES as well as LV ejection fraction, and a range of Doppler echocardiography derived biomarkers.

However, despite the richness of available data from 2-D echocardiography, which typically consists of a cine sequence of images covering one or more complete heart beats, current methods have mostly focused on estimating functional biomarkers from only 2 frames of this sequence (ED and ES). Even when the complete sequence was used as input to the model \cite{ouyang2020video}, the biomarkers estimated (e.g. LV ejection fraction) were limited to those based on only these two frames. Clinically speaking, a much richer description of cardiac function can be obtained by estimating functional biomarkers from the complete cardiac cycle, resulting in biomarkers such as ejection and filling rates and first-phase ejection fraction. These have been shown to be valuable biomarkers for earlier detection and monitoring of disease \cite{gu2019first,rokey1985determination}. In cardiac magnetic resonance (CMR) imaging, AI techniques have been proposed for automatically estimating this richer set of biomarkers \cite{Ruijsink2020}. However, estimating them from echocardiography has historically required a significant amount of expert interaction, i.e. for contouring the boundaries of the LV myocardium over the entire cardiac cycle. 

In this work we present, for the first time, an AI approach for estimating a much richer set of cardiac functional biomarkers from 2-D echocardiography. The biomarkers describe both systolic and diastolic LV function, enabling us to paint a much richer picture of the heart in health and disease.

\section{Materials}
\label{sec:Materials}
Four datasets were used for the training and testing of our model for 2-D echocardiography image analysis, and
these are described below:\\
\textbf{1. EchoNet-Dynamic \cite{ouyang2020video}:} Apical 4-chamber (4Ch) echocardiography images were acquired by skilled sonographers using IE33, Sonos, Acuson SC2000, Epiq 5G, or Epiq 7C ultrasound machines on a cohort of 10,030 patients, split into 8,742 and 1,288 patients respectively, for the training and test sets. The resolution of the echocardiography images was 600 $\times$ 600 or 768 $\times$ 768 pixels depending on the ultrasound machine and the images were downsampled by cubic interpolation using OpenCV into standardized 112 $\times$ 112 pixel videos. For all subjects, the LV endocardial border was manually traced in the ED and ES frames. For further details of the image acquisition protocol, see \cite{ouyang2020video}. \\
\textbf{2. Cardiac Acquisitions for Multi-structure Ultrasound Segmentation (CAMUS) \cite{leclerc2019deep}:} The CAMUS dataset consists of clinical exams from 450 patients acquired using GE Vivid E95 ultrasound scanners (GE Vingmed Ultrasound, Horten Norway) with a GE M5S probe (GE Healthcare, US). This database was split into 400 and 50 patients respectively, for the training and test sets. Three cardiologists manually delineated the LV endocardial, epicardial and LA borders in the ED and ES frames using a semi-automatic tool. For further details of the image acquisition protocol, see \cite{leclerc2019deep}. \\
\textbf{3. GSTFT echocardiography database:} This database contains 4Ch echocardiography data acquired from a cohort of 62 heart failure with reduced ejection fraction (HFrEF) patients imaged at Guy's and St. Thomas' Foundation Trust (GSTFT). The study was approved by the London Research Ethics Committee (11/LO/1232), all patients provided written informed consent for participation and the research was conducted according to the Helsinki Declaration guidelines on human research. This database was split into 50 and 12 subjects respectively for training and test. The ultrasound machines used were Philips IE33 and EPIQ 7C (Phillips Medical Systems, Andover, MA, USA) and the GE Vivid E9 (General Electric Medical Health, Milwaukee, WI, USA), each equipped with a matrix array transducer. The echocardiography images had an in-plane resolution between 0.26x0.26mm\textsuperscript{2} and 0.62×0.62mm\textsuperscript{2}. For all subjects, the endocardial borders and the ultrasound cone were manually traced in the ED and ES frames, and for the training database, four other random frames in the cardiac cycle were manually segmented.\\
\textbf{4. GSTFT paired echocardiography and CMR database:} This database contains paired CMR and 4Ch echocardiography data for a cohort of 50 heart failure patients acquired at GSTFT. This dataset was used only for testing the proposed framework. The dataset has similar echocardiography image parameters and patient characteristics to the GSTFT echocardiography database described above. CMR imaging was carried out on multiple scanners: Siemens Aera 1.5T, Siemens Biograph mMR 3T, Philips 1.5T Ingenia and Philips 1.5T and 3T Achieva, and the cine short axis sequences had a slice thickness between 8 and 10 mm and an in-plane resolution between 0.92x0.92mm\textsuperscript{2} and 2.4×2.4mm\textsuperscript{2}. 

\section{Methods}
\label{sec:methods}
\begin{figure}[!t]
\centering
\includegraphics[width=\textwidth]{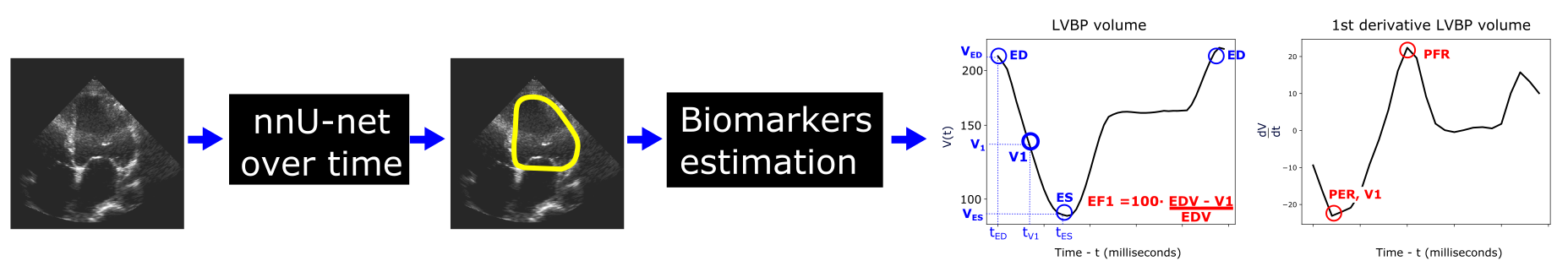}
\caption{Overview of the proposed framework to segment the left ventricle blood pool (LVBP) from echocardiography apical 4Ch images and to derive the systolic and diastolic clinical biomarkers.}
\label{fig:overview}
\end{figure}

An overview of the proposed framework is shown in Fig. \ref{fig:overview}. First, the trained DL-based echocardiography segmentation model is applied to produce left ventricle blood pool (LVBP) segmentations throughout the cardiac cycle. Next, the LVBP volume curve is computed and finally the advanced biomarkers of systolic and diastolic function are computed.

\noindent \textbf{Pre-processing:} For the GSTFT echocardiography database, an automated pre-processing workflow was developed to remove identifying information and eliminate any manual annotations on the images.
First, a `nnU-Net' segmentation network \cite{ouyang2020video} was trained using the GSTFT database to detect the ultrasound cone and remove any image information outside this cone. Second, we used an inpainting and multiplicative denoising algorithm \cite{barnes2009patchmatch} to remove any manual annotations present inside the image cone. \\
\noindent \textbf{Echocardiography apical 4Ch segmentation:} We used the `nnU-Net' network \cite{isensee2021nnu} for automatic segmentation of the LVBP from  apical 4Ch echocardiography images. The model was pre-trained using the training sets of the EchoNet-Dynamic database \cite{ouyang2020video} and the CAMUS database \cite{leclerc2019deep} with annotations at ED and ES frames. To take into account the inter-vendor differences in intensity distributions, the segmentation model was then fine-tuned using 300 GSTFT images. At test time, the test sets of the GSTFT,  EchoNet-Dynamic and CAMUS databases were used for inference.

The nnU-Net segmentation network was trained and evaluated using a five-fold cross validation on the training set. As in \cite{isensee2021nnu}, the network was trained for 1,000 epochs, where one epoch is defined as an iteration over 250 mini-batches (with a batch size of 30). Stochastic gradient descent with Nesterov momentum ($\mu$=0.99) and an initial learning rate of 0.01 was used for learning network weights. The loss function used to train the `nnU-Net' model was the sum of cross entropy and Dice loss. Data augmentation was performed on the fly and included techniques such as rotations, scaling, Gaussian noise, Gaussian blur, brightness, contrast, simulation of low resolution, gamma correction and mirroring. Please refer to \cite{isensee2021nnu} for more details of the network training.\\
\noindent \textbf{Biomarker estimation:} LVBP volume curves were calculated from the obtained segmentations using the single-plane modified Simpson’s method of discs \cite{folland1979assessment}.
From the LVBP volume curves, the following clinical biomarkers were computed:
\indent \textit{Standard clinical biomarkers}: LVBP end-diastolic volume (EDV), end-systolic volume (ESV) and ejection fraction (EF).\\
\indent \textit{Advanced systolic/diastolic biomarkers}: LVBP first-phase ejection fraction (EF1), LVBP peak ejection rate (PER), LVBP peak filling rate (PFR).\\
Below we refer to the LVBP volume curves as $V(t)$, with $t$ representing the acquisition time in milliseconds. Please refer to Fig. \ref{fig:overview} for an illustration of the following definitions. 

The ED and ES frames were determined as the frames with maximum and minimum LVBP volume respectively, and the EDV and ESV correspond to the volumes at these frames.
Based upon these, EF was computed using the standard formula: $EF = 100 \times (EDV - ESV) / EDV$.
EF1 can be defined as the percentage change in LVBP volume from ED to the time of peak aortic valve (AV) flow \cite{gu2019first,GU20212275}. In the absence of continuous wave Doppler AV flow images, the time of peak AV flow is assumed to be equivalent to the time of the minimum first derivative of LVBP volume in systole \cite{gu2019first,GU20212275}, i.e.
\begin{eqnarray}
EF1 = & 100 \times (EDV - V1) / EDV, \\
\mbox{where~} V1 = & V(t_{V1}) \nonumber \\
\mbox{~and~} t_{V1} = & \argmin_t dV/dt, t \in [t_{ED}, \ldots t_{ES}] \nonumber
\end{eqnarray}
PER measures the slope of the ejection phase of the cardiac cycle. PER can be computed as follows \cite{bacharach1979left}:
\begin{equation}
PER = \Delta_t \times \min(dV/dt), t \in [t_{ED},...t_{ES}]
\end{equation}
where $\Delta_t$ represents the time in milliseconds between consecutive echocardiography frames.

PFR measures the slope of the rapid inflow phase of the cardiac cycle. PFR can be computed as follows \cite{bacharach1979left}:
\begin{equation}
PFR = \Delta_t \times \max(dV/dt), t \in [t_{ES}, \ldots t_{ED}]
\end{equation}

\noindent \textbf{Evaluation:} For quantitative assessment of the image segmentation model, we used the Dice similarity coefficient (DSC), which quantifies the overlap between an automated segmentation and a ground truth segmentation. We evaluated the accuracy of the derived clinical biomarkers by computing the absolute differences between automated and manual measurements, and by Bland-Altman analysis and Pearson’s correlation analysis. To verify the significance of the biases, paired t-tests versus zero values were applied.

\section{Experiments}
\label{sec:results} 

The proposed framework automatically generates segmentations of the LVBP over the full cardiac cycle and estimates the standard and advanced systolic/diastolic clinical biomarkers. Three sets of experiments were performed to evaluate the framework's performance. The first set of experiments aimed to validate the segmentation model in terms of overlap performance and clinical biomarker estimation, the second set of experiments aimed to evaluate the utility of the biomarkers for patient stratification, while the third set of experiments aimed to compare the biomarkers derived from CMR and echocardiography data. Our implementation is publicly available at \url{https://github.com/estherpuyol/ASMUS\_2022.git}.

\begin{figure}[h]
\centering
\includegraphics[width=\textwidth]{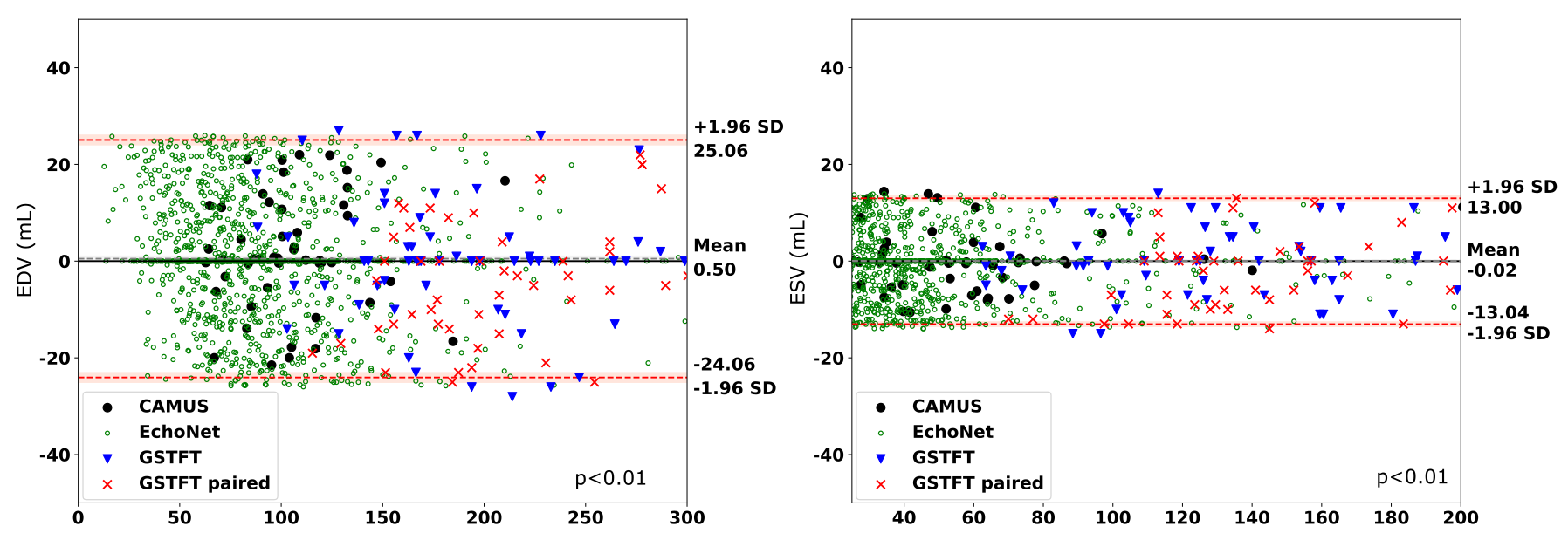}
\caption{Bland-Altman plots for the standard clinical biomarkers: Left ventricular end-diastolic volume (EDV) and left ventricular end-systolic volume (ESV); the red dotted lines show the limits of agreement. The p values represent the difference in mean bias from zero using a paired t-test. }
\label{fig:BA}
\end{figure}

\begin{figure}[h]
\centering
\includegraphics[width=\textwidth]{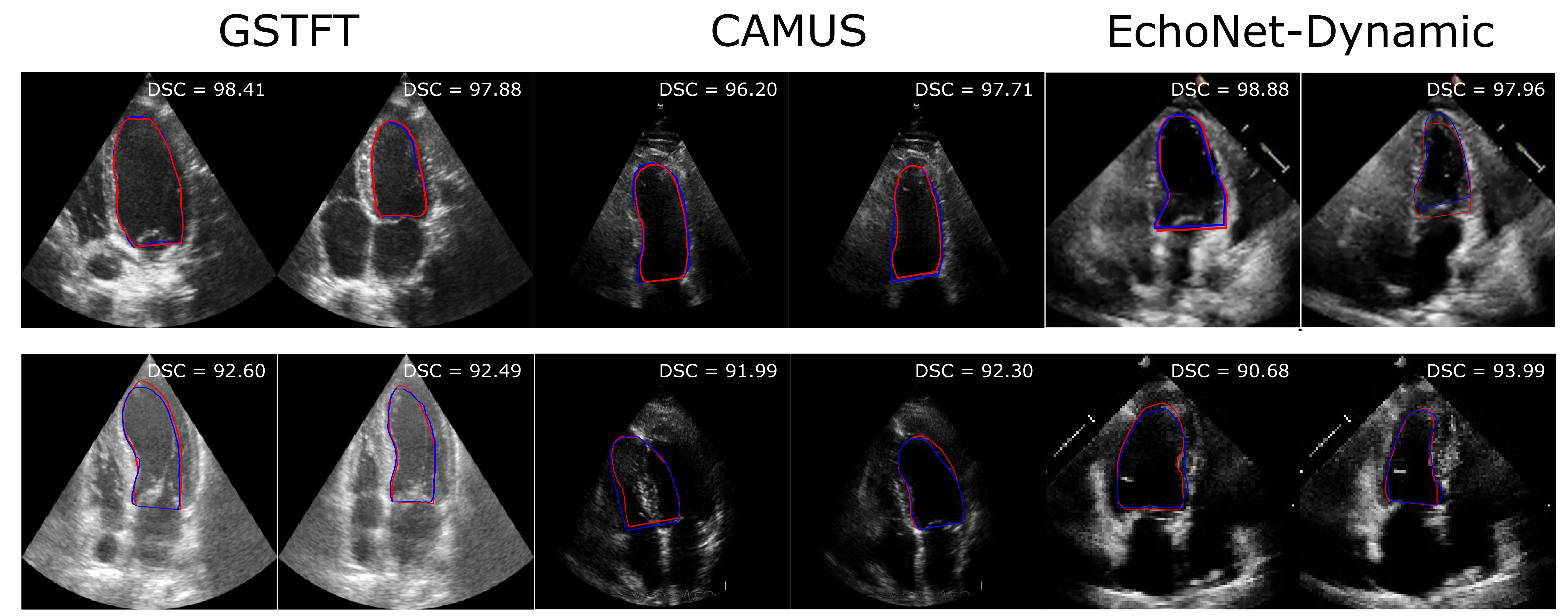}
\caption{Qualitative visual assessment of automated segmentation. Illustration of the segmentation results for each database for cases with high (top row) and low (bottom row) DSC.}
\label{fig:examples_Echo}
\end{figure}

\noindent \textbf{Experiment 1 - Comparative evaluation of the automated segmentation network:}
The DSC between automated and manual segmentations (only ED and ES frames, for which ground truths were available) for the test sets of all four databases (databases 1-4 in Section \ref{sec:Materials}) are summarised in the first block of Table \ref{table:1}. These results are in line with previous published methods \cite{ouyang2020video,leclerc2019deep}. The second block of Table \ref{table:1} shows the difference in standard clinical biomarker values between automated segmentation and manual segmentation. Figure \ref{fig:BA} shows the Bland-Altman plots for agreement between the pipeline and manual analysis, and Figure \ref{fig:examples_Echo} shows a visual comparison between the automated and manual segmentations for a set of test cases with high and low DSC.
Note that this experiment only focuses on the standard clinical biomarkers and not the advanced systolic/diastolic biomarkers because of the lack of ground truth segmentations at cardiac phases other than ED and ES.

~

\begin{table} [h]
\centering
\caption{Comparative evaluation of the automated segmentation network. The first block shows the Dice similarity coefficient (DSC) between automated segmentation and manual segmentation for the four test sets, and the second block shows the absolute differences for the standard clinical biomarkers for the same test sets. All metrics are expressed as mean (STD).}
\begin{tabular}{ L{3.6cm} C{3.2cm} C{3.2cm}} \hline 
\multicolumn{3}{c}{\textbf{DSC - Dice coefficient}}\\
Database & ED & ES \\ \hline 
1. EchoNet-Dynamic & 0.935 (0.03) & 0.926 (0.03)\\
2. CAMUS  & 0.931 (0.07) & 0.922 (0.06)\\ 
3. GSTFT & 0.928 (0.03) & 0.919 (0.04)\\
4. GSTFT paired & 0.920 (0.04) & 0.916 (0.04)\\ \hline \hline
\end{tabular}
\begin{tabular}{ L{3.1cm} C{2.3cm} C{2.3cm} C{2.3cm}} 
\multicolumn{4}{c}{\textbf{Absolute difference between auto and manual}}\\
Database & EDV (mL) & ESV (mL) & EF (\%) \\ \hline
1. EchoNet-Dynamic & 8.9 (8.8) & 4.7 (4.8) & 1.4 (1.3)\\
2. CAMUS & 9.3 (7.8) & 4.9 (4.5) & 1.5 (1.4)\\ 
3. GSTFT & 10.2 (9.6) & 5.6 (4.5) & 1.9 (1.8)\\
4. GSTFT paired & 10.8 (7.4) & 6.4 (4.8) & 1.8 (1.2)\\ \hline
\end{tabular}
\label{table:1}
\end{table}

\begin{figure}[h]
\centering
\includegraphics[width=\textwidth]{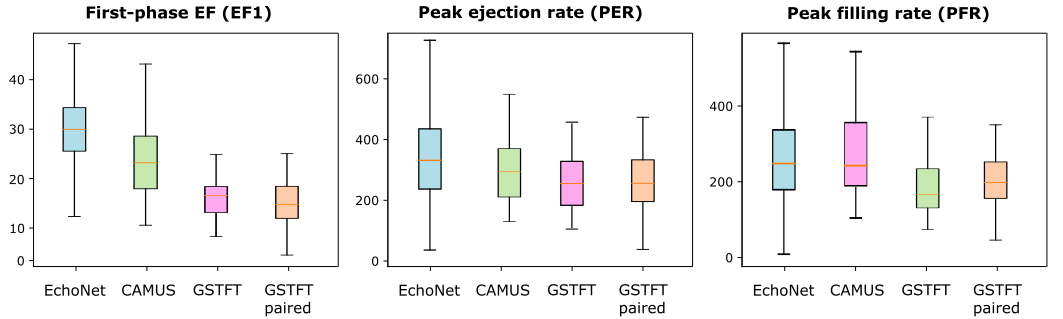}
\caption{Comparison of the advanced systolic/diastolic biomarkers for four test databases. Blue corresponds to EchoNet, green to CAMUS and pink to GSTFT.}
\label{fig:boxplots}
\end{figure}

\noindent \textbf{Experiment 2 - Analysis of biomarkers for patient stratification:} In this experiment we investigated the distribution of the clinical biomarkers for the four different databases (databases 1-4 in Section \ref{sec:Materials}).
Table \ref{table:volumes} shows the mean and standard deviation (STD) for all the derived biomarkers, both standard and advanced, and Figure \ref{fig:boxplots} shows boxplots for only the advanced systolic/diastolic biomarkers. The results show the ability of the proposed framework to robustly derive advanced biomarkers of cardiac function for different databases with significantly different populations. For example, the CAMUS and EchoNet-Dynamic databases have mixed populations, with mainly healthy cases and cases with mild HFrEF, while the GSTFT databases are composed mainly of patients with severe HFrEF. Some differences between these groups can be observed in the standard biomarkers, for example ejection fraction is lower in the GSTFT group compared to the other groups. However, these differences are more marked in the advanced systolic/diastolic biomarkers.

~

\begin{table} [h]
\centering
\caption{Comparison between the clinical biomarkers on four different test databases.  End-diastolic volume (EDV),  end-systolic volume (ESV), left ventricular ejection fraction (EF), first-phase ejection fraction (EF1), peak ejection rate (PER) and peak filling rate (PFR). All metrics are expressed as mean (STD).}
\begin{tabular}{l c c c c c c } \hline 
\multicolumn{7}{c}{\textbf{Clinical biomarkers}}\\
Database & EDV (mL) & ESV (mL) & EF (\%) & EF1 (\%) & PER (mL/s) & PFR (mL/s) \\ \hline 
1. EchoNet-Dynamic & 92 (46) & 44 (36) & 55 (12) & 30 (8) & 350 (57) & 226 (117)\\
2. CAMUS  & 110 (41) & 56 (33) & 51 (13) & 24 (8) & 305 (108) & 270 (106)\\ 
3. GSTFT & 203 (98) & 139 (03) & 30 (10) & 16 (7) & 266 (95) & 197 (95)\\
4. GSTFT paired & 207 (55) & 148 (45) & 29 (7) & 16 (6) & 272 (108) & 209 (99)\\\hline 
\end{tabular}
\label{table:volumes}
\end{table}

\noindent \textbf{Experiment 3 - Comparison between CMR and echocardiography derived biomarkers:} The functional biomarkers derived from echocardiography were compared with corresponding values derived from CMR using Pearson’s correlation and limits of agreement (1.96xSD = 95\% confidence intervals) determined by Bland-Altman analysis \cite{bland1986statistical}. A paired Student's $t$-test was used to test for any overestimation or underestimation between echocardiography and CMR in the Bland-Altman analysis. As further validation, we compute the  root mean square error (RMSE) between the CMR and echocardiography biomarkers. Table \ref{table:comp_CMR_Echo} shows in the first two rows the biomarkers estimated from the GSTFT paired echocardiography and CMR database, and the third row the RMSE. The remaining rows show the results of the correlation analysis. The CMR database was analysed using a state-of-the-art automated cardiac functional quantification framework \cite{Ruijsink2020}. 
\begin{table} [h]
\centering
\caption{CMR and echocardiography derived biomarkers and correlation analysis.  End-diastolic volume (EDV),  end-systolic volume (ESV), left ventricular ejection fraction (EF), first-phase ejection fraction (EF1), peak ejection rate (PER) and peak filling rate (PFR). Correlation coefficient (R), root mean square error. p-value derived from the paired Student $t$-test. All clinical biomarkers are expressed as mean (STD). }
\begin{tabular}{l c c c c c c } \hline 
\multicolumn{7}{c}{\textbf{Derived clinical biomarkers}}\\
Database & EDV (mL) & ESV (mL) & EF (\%) & EF1 (\%) & PER (mL/s) & PFR (mL/s) \\ \hline 
Echocardiography & 207 (55) & 148 (45) & 29 (7) & 18 (6) & 272 (108) & 209 (99) \\
CMR & 212 (54) & 146 (42) & 31 (8) & 18 (6) & 367 (139) & 280 (127) \\ \hline
RMSE &  29 & 18 & 6 & 4 & 133 & 93  \\ \hline
\multicolumn{7}{c}{\textbf{Correlation Analysis between echocardiography and CMR }}\\
Parameter & EDV (mL) & ESV (mL) & EF (\%) & EF1 (\%) & PER (mL/s) & PFR (mL/s) \\ \hline 
R & 0.86 & 0.92 & 0.79 & 0.77 & 0.74 & 0.88 \\
p-value & 0.61 & 0.85 & 0.13 & 0.14 & 0.001 & 0.001 \\
\end{tabular}
\label{table:comp_CMR_Echo}
\end{table}

Correlation analysis showed a very strong relationship between CMR and echocardiography for all biomarkers \cite{dougan2018bland}. Bland-Altman analysis demonstrated no significant difference (p $>$ 0.05 for all biomarkers except PER and PFR) for the limits of agreement for echocardiography biomarkers in comparison with CMR biomarkers. For the standard clinical biomarkers, the results are in line with previous publications \cite{greupner2012head,rigolli2016bias}.

\section{Discussion}
\label{sec:discussion}

In this paper we have presented the first AI framework for automatically estimating a range of advanced biomarkers of cardiac function. Specifically, we estimate first phase ejection fraction (EF1), peak ejection rate (PER) and peak filling rate (PFR). These biomarkers are different to the standard functional biomarkers such as EDV, ESV and EF in that they make use of the dynamic change in ventricular volumes throughout the cardiac cycle rather than just two frames (ED and ES). As a result they have the potential to provide a more complete description of cardiac function and this has been shown in recent clinical studies \cite{gu2019first,rokey1985determination}. However, this also means that it is currently not feasible to estimate them in clinical practice due to the labour-intensive nature of the required manual segmentations. Our work has shown for the first time that these biomarkers can be automatically estimated from echocardiography images using deep learning.

Our proposed method achieves similar segmentation overlap performance to the original EchoNet-Dynamic model \cite{ouyang2020video}, e.g. mean ED DSC is 0.929 (95\% confidence interval 0.926–0.935) vs 0.927 (95\% confidence interval 0.925–0.928) in EchoNet-Dynamic. For the standard clinical biomarkers, our method's performance is in line with results reported in the EchoNet-Dynamic and CAMUS papers \cite{ouyang2020video,leclerc2019deep}, e.g. we achieved an absolute mean error for EDV of 8.9mL compared to 9.5ml reported in the CAMUS paper, and an absolute mean error for EF of 1.4\% compared to 4.22\% in the EchoNet-Dynamic paper.

We believe that this work is clinically important and timely. There is an increasing acceptance in the clinical community that biomarkers such as EDV, ESV and EF are not optimal for use in patient stratification in heart failure. Current clinical guidelines \cite{Ponikowski2016} indicate that heart failure patients should be stratified for treatment mainly using a single biomarker, EF. However, the most recently published guidelines recognise the limitations of over-reliance on only one measure of cardiac function. Using this stratification there is a wide range of response to treatment - the hazard ratio for mortality for pharmacological treatment for heart failure is 0.44, with a 95\% confidence interval of  0.26-0.66 \cite{Burnett2017}. This wide confidence interval highlights the need for improved patient stratification in heart failure and new advanced biomarkers of function are likely to play a significant role in this. As well as demonstrating the robustness and accuracy of our framework (Experiments 1 and 3), in this paper we have demonstrated how these advanced biomarkers can lead to better stratification in heart failure patients (Experiment 2).

This paper represents an important proof-of-concept, but it has its limitations. 
reasonable frame rate ($\geq$ 20 frames/second). A lower frame rate may cause difficulties in estimating the volume derivatives needed to compute the advanced systolic/diastolic biomarkers. Future work will investigate strategies for dealing with data with lower frame rates. In addition, in future work we plan to perform a more thorough validation of our framework using data with ground truth annotations throughout the cardiac cycle. This would enable us to compute errors in the biomarker estimates compared to ground truth values. We also plan to investigate a wider range of advanced biomarkers from echocardiography data and more thoroughly investigate their utility in heart failure patient stratification.

\section*{Acknowledgements}
This work was supported by the EPSRC (EP/R005516/1 and EP/P001009/1), the Wellcome EPSRC Centre for Medical Engineering at the School of Biomedical Engineering and Imaging Sciences, King\textquotesingle s College London (WT 203148/Z/16/Z). The authors acknowledge financial support (support) the National Institute for Health Research (NIHR) Cardiovascular MedTech Co-operative award to the Guy’s and St Thomas’ NHS Foundation Trust and the Department of Health National Institute for Health Research (NIHR) comprehensive Biomedical Research Centre award to Guy’s \& St Thomas’ NHS Foundation Trust in partnership with King’s College London

\bibliographystyle{splncs04}
\bibliography{refs}

\end{document}